# Attosecond vortex pulse trains

**ALBA DE LAS HERAS,**[1,2,4] **DAVID SCHMIDT,**[3,5] **JULIO SAN ROMÁN,**[1,2] **JAVIER SERRANO,**[1,2] **DANIEL ADAMS,**[3] **LUIS PLAJA,**[1,2] **CHARLES G. DURFEE,**[3] AND **CARLOS HERNÁNDEZ-GARCÍA**[1,2]

[1]*Grupo de Investigación en Aplicaciones del Láser y Fotónica, Departamento de Física Aplicada, Universidad de Salamanca, E-37008 Salamanca, Spain*
[2]*Unidad de Excelencia en Luz y Materia Estructuradas (LUMES), Universidad de Salamanca, Salamanca, Spain*
[3]*Department of Physics, Colorado School of Mines, Golden, Colorado 80401, USA*
[4]*albadelasheras@usal.es*
[5]*daschmid@mines.edu*

**Abstract:**

The landscape of ultrafast structured light pulses has recently evolved driven by the capability of high-order harmonic generation (HHG) to up-convert orbital angular momentum (OAM) from the infrared to the extreme-ultraviolet (EUV) spectral regime. Accordingly, HHG has been proven to produce EUV vortex pulses at the femtosecond timescale. Here we demonstrate the generation of attosecond vortex pulse trains, i.e. a succession of attosecond pulses with a helical wavefront, resulting from the synthesis of a comb of EUV high-order harmonics with the same OAM. By driving HHG with a polarization tilt-angle fork grating, two spatially separated circularly polarized high-order harmonic beams with order-independent OAM are created. Our work opens the route towards attosecond-resolved OAM light-matter interactions.

## 1.  Introduction

Reaching attosecond electromagnetic pulse durations [1,2] was a remarkable breakthrough that has provided access to observe and manipulate the ultrafast electron dynamics in atoms, ions, molecules, liquids, or solids [3–9]. The generation of attosecond pulses was possible thanks to the nonlinear optical phenomena of high-order harmonic generation (HHG) [10], where multiple harmonic frequencies of the fundamental are emitted in the interaction of an intense infrared (IR) femtosecond (fs) laser pulse with a noble gas target [11]. The spectrum resulting from HHG is characterized by a nonpertubative plateau of high-order harmonics that exhibit similar intensity [12], extending to the extreme ultraviolet (EUV) or even the soft X-ray regime [13]. One of the most relevant aspects of HHG is that it provides a comb of harmonics whose relative phase is regular [14–18]. As a consequence, the coherent superposition of the higher-order harmonics within the plateau gives rise to a train of attosecond pulses [1].

The atomic picture of HHG can be explained by a simple three-step model [19–21] occurring every half-cycle of the driving fs laser field. First, the outermost electron tunnels through the atomic Coulomb potential barrier distorted by the laser field; subsequently, the free electron acquires kinetic energy during the excursion in the continuum; and finally, it comes back to the parent ion, emitting a high-energy photon upon recollision. If the interaction is restricted to a single half-cycle event—for example, by using a few-cycle driving pulse—a supercontinuum spectrum is obtained, corresponding to the emission of an isolated attosecond pulse [2]. As soon as several events efficiently occur, a comb of harmonics is produced, resulting in an attosecond pulse train. Still, different techniques based on temporal gating or advantageous phase-matching configurations can also produce isolated attosecond pulses [8,22–30]. It is important to note that the efficiency of the third step—the recombination

step—, drastically decreases with the driving field ellipticity, as the electron trajectories do not return to the parent ion [21]. As a consequence, HHG was restricted for many years to produce linearly polarized harmonics [31,32] and attosecond pulses. Then, several ingenious techniques based on designing a proper driving field have provided access to the production of circularly polarized harmonics [33–39]. Among them, the combination of two counter-rotating circularly polarized fs driving pulses in a noncollinear scheme has allowed for the generation of circularly polarized attosecond pulses [40,41]. In such a scheme, the driving field at the focal plane is linearly polarized—ensuring efficient recombination upon HHG in the gas target— with the polarization tilt-angle rotating along the transverse axis. Upon propagation into the far field, two separate circularly polarized harmonic beams with opposite handedness are emitted, allowing for the synthesis of counter-rotating circularly polarized attosecond pulse trains [40], and circularly polarized isolated attosecond pulses—if driven by few-cycle fs pulses [41,42]. It is worth noting that when two linearly polarized beams are crossed at the gas jet, an angular array of harmonic beams is emitted, in directions satisfying the conservation of linear momentum [43,44]. By using two counter-rotating circularly polarized pulses, conservation of spin angular momentum (SAM) restricts the process to only produce two harmonic beams [40].

Aside from controlling the polarization state or SAM of the attosecond pulses, imprinting orbital angular momentum (OAM) would enable new possibilities to explore ultrafast electron dynamics in laser-matter interactions. Light beams carrying OAM—known as vortex beams—present a helical wavefront associated with their distinctive phase distribution along the transverse azimuthal coordinate [45,46]. They are characterized by their topological charge, $\ell$, which denotes the number of $2\pi$ phase shifts along the azimuthal coordinate. The corresponding OAM of the beam is $\hbar\ell$ [45,46], with $\hbar$ being the reduced Planck constant. The transfer of OAM from IR vortex beams to valence electrons [47,48] and photoelectrons [49] has already been experimentally demonstrated. Still, the higher frequency EUV or even X-ray attosecond regime is yet to be investigated, where theoretical studies predict unique signatures of the light's OAM and SAM in photoionization [50,51].

During the last decade, HHG has emerged as a unique mechanism to provide EUV harmonic vortex beams by nonlinearly up-converting IR laser beams carrying OAM [52–70]. By properly modifying the driving pulse, it has been demonstrated the generation of linearly polarized harmonic vortices in collinear [52,53,63–65,67,68] and noncollinear geometries [58–60], circularly polarized harmonic vortices [54,55,70], EUV beams with time-dependent OAM (i.e. self-torque) [66], or vector-vortex EUV harmonics [69], among others. When a driving pulse with OAM of $\hbar\ell$ is converted to the $q^{th}$-order harmonic, the resultant harmonic carries OAM of $\hbar q\ell$. This fundamental law imposes limitations on the generation of attosecond vortex pulses, since it inherently hinders the obtention of several harmonic orders with the same topological charge (q-independent OAM). This requirement is essential for creating an attosecond pulse with a consistent OAM. Consequently, to the best of our knowledge, the synthesis of attosecond vortex pulses has remained an outstanding challenge, primarily due to the absence of schemes enabling the generation of multiple harmonic vortex beams exhibiting the same topological charge and spatially overlapping.

In this article, we introduce the generation of attosecond vortex pulse trains, a set of attosecond bursts whose electric field helically twists according to a well-defined OAM. We demonstrate theoretically and experimentally that HHG driven by a polarization tilt-angle fork grating results in two isolated far-field EUV harmonic beams with opposite handedness. In this configuration, all the harmonics within each beam exhibit the same topological charge. This outcome arises due to the simultaneous conservation of linear momentum, SAM, and OAM. Then, by filtering the high harmonic orders, circularly polarized attosecond vortex pulse trains can be synthesized. Our result provides a unique tool to explore ultrafast light-matter interactions including SAM-and-OAM-driven electron dynamics. One could foresee applications in attosecond spectroscopy to probe chiral molecules, magnetic media, or biological samples through SAM-OAM-induced dichroism. Indeed, circular or helical dichroism that relies on selective SAM or OAM helicity absorption, respectively, has been successfully applied in different experiments [39,71–73].

## 2. The concept of attosecond vortex pulse trains

An attosecond vortex pulse train is a structured light beam composed of laser pulses of attosecond duration exhibiting a helical wavefront. It corresponds to the coherent superposition of several high-order harmonic beams carrying the same topological charge, $\ell$. Thus, we can characterize an attosecond vortex pulse train through the topological charge of their harmonic components. Illustrations of attosecond vortex pulse trains with $\ell = 1$ and $\ell = 2$ are shown in Figs. 1a and 1b, respectively. They correspond to the synthesis of five standard odd harmonic orders (from the 17$^{th}$ to the 25$^{th}$, as depicted in the upper right inset spectra), of a fundamental frequency $\omega_0$. The isosurface plots the structure of the electric field, which exhibits three attosecond bursts with a helical wavefront associated to $\ell$. For $\ell = 1$, the electric field describes a single spatiotemporal helicoid; whereas $\ell = 2$ implies two intertwined helicoids. Note that as $\ell$ can take any integer value, the number of helicoids (and their handedness) can be easily tuned. In addition, the spatially integrated intensity (filled light blue lines) reveals the attosecond temporal duration of each of the vortex pulses within the train (considering a standard central wavelength of 800 nm).

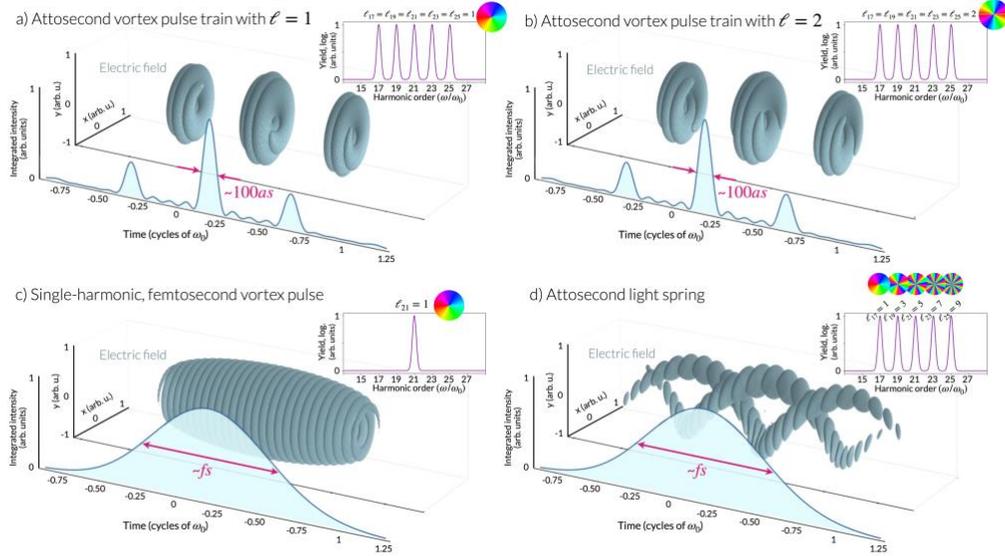

Fig. 1. Illustration of an attosecond vortex pulse train with a) $\ell = 1$ and b) $\ell = 2$. It corresponds to five standard harmonic orders (the 17$^{th}$ to the 25$^{th}$, shown in the upper right inset), carrying each of them a topological charge of $\ell = 1$ and $\ell = 2$ respectively. The isosurface encloses the spatiotemporal distribution of a positive value of the electric field, whereas the front light blue-filled two-dimensional plot depicts the spatially integrated intensity. For comparison purposes, c) depicts the fs vortex pulse associated with a single harmonic (the 21$^{st}$) with $\ell = 1$; and d) an attosecond light spring associated with five harmonic orders, from the 17$^{th}$ to the 25$^{th}$, with a topological charge that scales with the harmonic order from 1 to 9. The spatial profile of the harmonics is modeled with standard Laguerre Gaussian modes at the focal plane, and, for simplicity, the harmonics are considered to be phase-locked, i.e. their intrinsic HHG phase is neglected.

It is important to distinguish this result from those found in previous works of OAM-HHG. As mentioned above, in conventional HHG driven by an IR vortex beam with topological charge $\ell$, OAM conservation leads to the generation of high-order harmonic vortices whose topological charge scales linearly with the harmonic order, i.e. $\ell_q = q\ell$ [53,58,64,65,67,68]. In this scenario, by filtering a single harmonic, an EUV fs vortex pulse emerges in the temporal domain. This is illustrated in Fig. 1c, where for simplicity we have considered the 21$^{st}$ harmonic carrying $\ell = 1$. If, on the other hand, several high-order harmonics—which exhibit different topological charges— are considered, an attosecond light spring is obtained. The superposition of different harmonics carrying a linearly scaling topological charge gives rise to an intensity profile describing a spatiotemporal helicoid, which was denoted as a light spring [74]. Attosecond light springs are thus naturally obtained when driving HHG with a single-OAM beam, as theoretically predicted in [53] and experimentally measured in [64]. Note that they have also been referred to as attosecond vortices because indeed they carry OAM, and can be interpreted as an attosecond pulse train that is delayed along the azimuthal coordinate. In Fig. 1d we illustrate an attosecond light spring composed of the same five harmonic orders as in Figs. 1a and 1b, but with topological charges that scale with the harmonic order from 1 to 9 (the selection of lower topological charges than those obtained in a single-OAM HHG experiment is for illustration purposes). Whereas the real part of the electric field follows two spatiotemporal intertwined helices, the spatially integrated intensity reveals the fs timescale duration of the beam.

## 3. Set-up: polarization tilt-angle fork grating

The key ingredient to generate an attosecond vortex pulse train is to obtain high-order harmonics with the same topological charge, that also present a fair spatial overlap and that can be refocused for experimental applications. In this work, we demonstrate that this can be achieved if the driving field is configured as a polarization tilt-angle fork grating at the gas target plane. This highly structured beam consists of a linearly polarized electric field, whose tilt-angle varies in the transverse plane forming a fork grating (Fig. 2b). In order to create a polarization tilt-angle fork grating, we arrange a noncollinear crossing of two circularly polarized IR vortex beams that have opposite helicity in both SAM and OAM, i.e. $\sigma_1 = -\sigma_2$ and $\ell_1 = -\ell_2$. Figure 2 depicts a schematic illustration of the noncollinear scheme used to create a polarization tilt-angle fork grating. Note that if the local intensities of the two beams are matched, the superposition of the two counter-rotating driving beams at the generation plane of the gas target gives rise to a linearly polarized electric field, thus maximizing the efficiency in single-atom HHG. The polarization tilt-angle fork grating results from the particular SAM-OAM combination of the driving fields, and it is transferred to all the generated harmonics.

In our experiment and theoretical simulation, we design a polarization tilt-angle fork grating by combining two counter-rotating circularly polarized beams with topological charges $\ell_{RCP} = -1$ and $\ell_{LCP} = 1$, as depicted in Fig. 2. We consider beam waists of 50 $\mu$m, 800 nm central wavelength, and a half crossing angle of $\theta_c = 0.85°$. The pulse duration is approximately 50 fs in the experiment, whereas a shorter envelope of 7.7 fs at full-width duration at half-maximum intensity is chosen to reduce the computational cost. The two beams are focused near the tip of an Argon gas jet with a cross-section opening of 300 $\mu$m and an axial length of 30 $\mu$m. In the Methods section, we extend the details about the experimental configuration, as well as those of the macroscopic HHG simulations. The resulting theoretical and experimental intensity profiles of the IR driving beam at the focal plane (where the polarization tilt-angle fork grating is formed) are presented in Fig. 2c. We note that in the experimental intensity profile in Fig. 2c, the forked structure can be seen owing to slight departures from ellipticity, along with the projection of the beams due to the small crossing angle. Still, it also demonstrates the presence of the forked polarization tilt-angle grating.

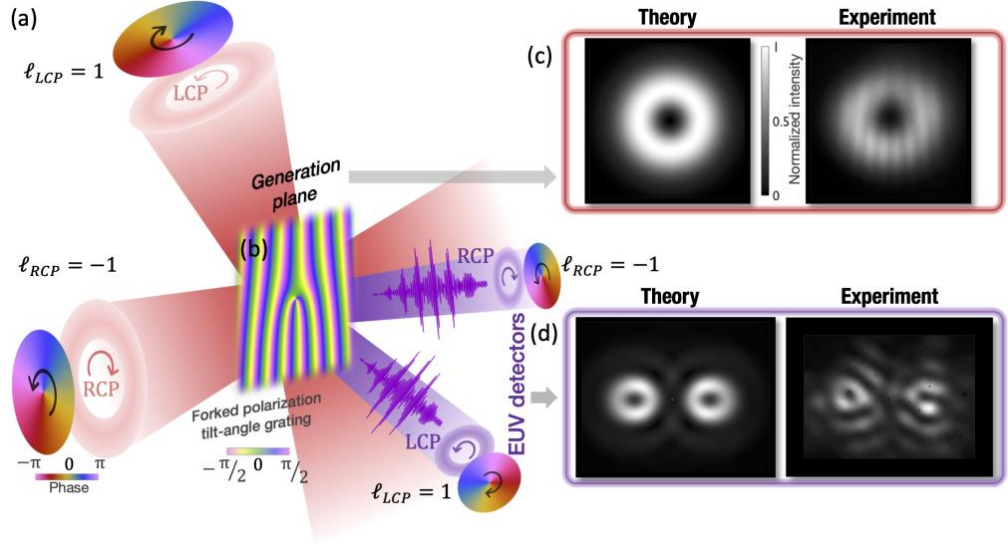

Fig. 2. Polarization tilt-angle fork grating scheme to generate attosecond vortex pulse trains. (a) Illustration of the noncollinear scheme, driven by two IR crossing beams that exhibit both opposite SAM and OAM. (b) Theoretical fork grating structure in the linear polarization orientation angle at the generation plane. Upon HHG, two separated harmonic beams are emitted in the form of attosecond vortex pulse trains. (c) Driving IR intensity profile at the generation plane in theory and experiment. (d) EUV far-field intensity profile resulting from the superposition of the $17^{th}$-$25^{th}$ harmonic orders.

## 4. Results

Figure 2d depicts the theoretical simulations and experimental results of the spatially resolved full harmonic beam profile detected in the far field. Once the harmonics are generated through the polarization tilt-angle fork grating at the gas jet, they propagate into the far field as two well-defined separated harmonic beams. The well-defined structure in the observed beams also demonstrates a satisfactory overlap of the detected $17^{th}$ to $25^{th}$ harmonic orders.

A more complete analysis of the harmonic emission is shown in Fig. 3. First, theoretical simulations in Fig. 3a show the spatially resolved RCP and LCP harmonic intensity, whereas in Fig. 3b, we depict their spatially resolved phase. The high-order harmonic emission is composed of two counter-rotating harmonic beams, with similar divergence, and with a very regular spatial phase distribution within each harmonic. In Fig. 3b we can already identify that all RCP (LCP) harmonics are emitted with a $2\pi$ phase azimuthal twist, thus carrying a topological charge of $\ell_{RCP} = -1$ ($\ell_{LCP} = 1$). This is corroborated in Fig. 3c, where we extract the OAM content of each harmonic by performing the Fourier transform of the numerical harmonic field along the azimuthal coordinate [75].

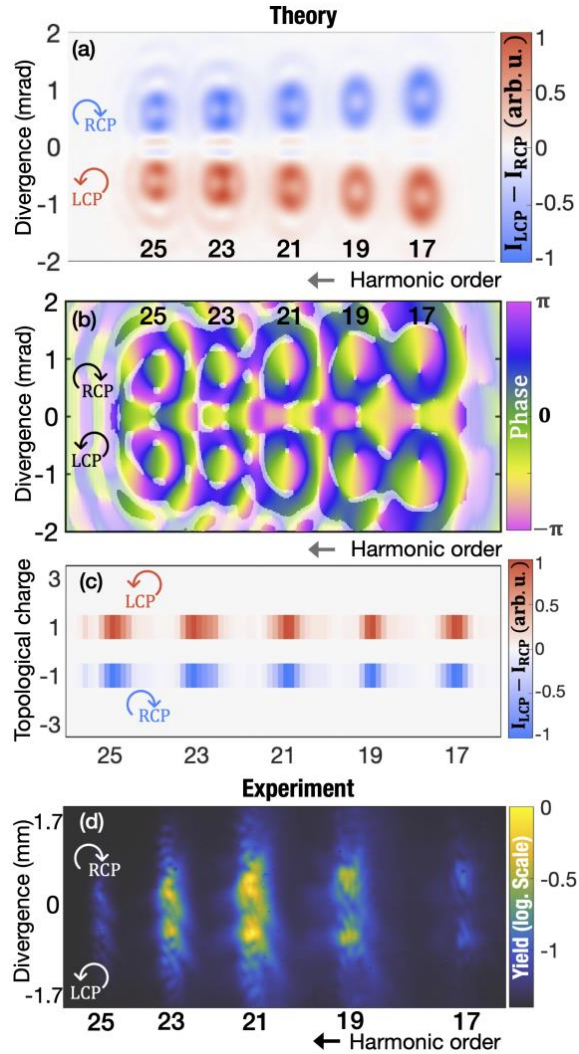

Fig. 3. Spectrally and spatially resolved (a) intensity and (b) phase profiles for the RCP and LCP harmonic components. (c) Topological charge content of the harmonic beams after performing the azimuthal Fourier Transform over the theoretical simulation results. (d) Spectrally and spatially resolved characterization of the high harmonic output from the experiment, showing the generation of two well-defined harmonic beams with a fairly good overlap of the harmonics.

The properties of the resulting harmonic beams can be understood through the simultaneous conservation of linear momentum, SAM, and OAM. At the photon picture, the $q^{th}$ harmonic order can be considered to be composed of $n_{RCP}$ photons from the RCP driver and $n_{LCP}$ from the LCP driver, satisfying $q = n_{RCP} + n_{LCP}$. SAM conservation establishes that $n_{RCP}$ and $n_{LCP}$ can only differ in one, i.e. $n_{RCP} = n_{LCP} \pm 1$ [40]. Together with linear momentum conservation, $\vec{k}_q = n_{RCP}\vec{k}_{RCP} + n_{LCP}\vec{k}_{LCP}$, being $\vec{k}_{RCP}$ and $\vec{k}_{LCP}$ the propagation wavevectors of the drivers, the far-field spatial position of each harmonic beam is found at a divergence angle of $\theta_q = \arctan(q^{-1}\tan\theta_c)$ [40]. Finally, applying the well-known OAM conservation law for two driving vortex beams [63], $\ell_q = n_{RCP}\ell_{RCP} + n_{LCP}\ell_{LCP}$, it is straightforward to demonstrate that all RCP (LCP) harmonics exhibit the same topological charge as the RCP

(LCP) driving field, i.e $\ell_{q,RCP} = \ell_{RCP} = -1$ ( $\ell_{q,LCP} = \ell_{LCP} = 1$ ). This simultaneous conservation of linear and angular momentum has been paramount in several previous works of HHG driven by structured beams [40,58–60,69,70].

The spatial and spectrally resolved experimental harmonic profile is shown in Fig. 3d. Note that the spectral dispersion does not allow to resolve the annular intensity distributions of the harmonics, but it can be inferred from the lower yield of the harmonic profiles at their center and from Fig. 2d. Most importantly, we observe, as in theory, that two separate beams of spatially overlapping high-order harmonics are produced in the far field. Thus, the two high-order harmonic beams presented in Fig. 3 exhibit all the ingredients to be emitted as two attosecond vortex pulse trains.

In Fig. 4 we present the spatiotemporal distribution of the electric field (top row) and intensity (bottom row) obtained after performing the Fourier transform of the simulated harmonics presented in Fig. 3. The well-defined OAM of each harmonic beam along all the spectrum, together with the good overlapping between the harmonics, results in the synthesis of two circularly polarized trains of attosecond pulses exhibiting a helically twisting electric field of opposite handedness. The comparison of the simulation results in Fig. 4a with Fig. 1a demonstrates that the proposed noncollinear scheme enables the generation of two separated attosecond vortex pulse trains. Whereas the electric field of each pulse exhibits the helical wavefront according to the OAM of the harmonic beams (Fig. 4a), the spatiotemporal intensity distribution, shown in Fig. 4b, presents a toroidal shape containing only the signature of a phase singularity at the center. To have a better insight into the temporal duration of each pulse within the train, in Figs. 4b,c and 4e,f we depict the real part and the intensity of the attosecond pulse train at the spatial positions where the RCP and LCP harmonic components are maximized. Attosecond vortex pulses of ~270 attoseconds duration full-width-at-half-maximum (FWHM) in intensity are obtained.

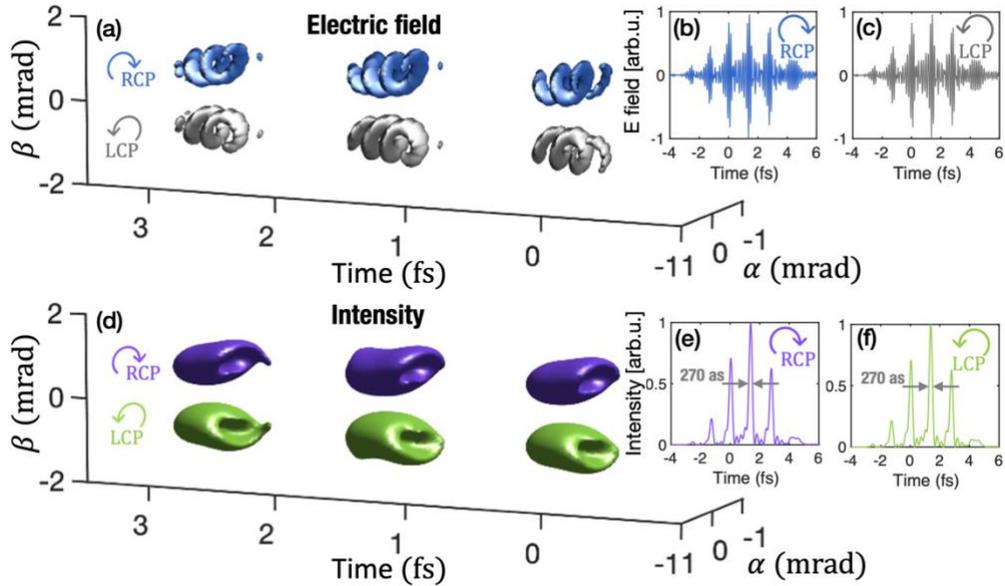

Fig. 4. (a) Spatiotemporal structure of the electric field obtained from the numerical simulations, showing the generation of two separated attosecond vortex pulse trains with opposite handedness in SAM and OAM. Panels (b) and (c) show the real part of the RCP and LCP attosecond pulse trains at the spatial position of maximum intensity. (d) Spatiotemporal structure of the intensity obtained from the numerical simulations. (e, f) Intensity of the RCP and

LCP attosecond pulse trains at the spatial position of maximum intensity. The isosurfaces of the real part and intensity correspond to a value that is 15% of the peak intensity.

It is important to highlight that the two RCP/LCP attosecond vortex pulse trains are well isolated in the far field, so a single train could be spatially filtered and refocused for applications. In the Supplemental Information, we provide simulation results for higher-order attosecond vortex pulse trains obtained by modifying the topological charge of the driving beams.

## 5. Discussion and Conclusions

In summary, we have introduced attosecond vortex pulse trains as novel ultrafast light structures exhibiting a well-defined OAM in the attosecond regime. They correspond to the synthesis of a high-frequency vortex comb where all the harmonic components exhibit the same topological charge. We have demonstrated that HHG driven by a structured beam that exhibits a polarization tilt-angle fork grating at the focal plane naturally results in two separated attosecond vortex pulse trains with opposite SAM and OAM helicities. Upon spatial filtering, a single attosecond vortex pulse train can be refocused to be used in experiments. We have provided theoretical and experimental evidence of the generation of this unique ultrafast light structure. Still, a complete experimental characterization of an attosecond vortex pulse train would require spatially-resolved EUV ultrafast temporal characterization techniques, which, to the best of our knowledge, are beyond the state-of-the-art technologies.

We emphasize that our work demonstrates the generation of attosecond vortex pulses with circular polarization states. This is a consequence of the favorable simultaneous conservation of linear and angular momentum, allowing for the generation of high-order harmonics with the same topological charge and, thus, circumventing the well-known linear scaling law for the OAM in HHG [53,58,64,67]. Previous works that have demonstrated tunability over the topological charge of the high-order harmonics in noncollinear geometries [58–60] could motivate the generation of linearly polarized attosecond vortex pulse trains upon optimizing the superposition of the different high-order harmonics with the same topological charge.

We also note that the generation of a train of attosecond vortex pulses is a direct consequence of the multi-cycle behavior of the driving pulses used. The isolation of a single vortex pulse is limited by the generation of few-cycle vortex driving pulses, or by the development of gating techniques that can support the favorable linear and angular momentum conservation laws for the generation of a supercontinuum with a single topological charge.

Advancing the spatiotemporal characterization techniques of attosecond pulses is paramount for obtaining further experimental evidence on the generation of pulses spatially structured in the attosecond regime. Then, by incorporating attosecond vortex pulses in spectroscopic measurements, we add an extra degree of freedom in the interaction with matter. The combination of attosecond pulses carrying both SAM and OAM can be particularly interesting in probing chirality and spin interactions, with applications in diverse scientific areas such as chemistry, biology, condensed matter physics, or magnetism.

## 6. Methods

### 6.1 Experimental methods

In the early stages of noncollinear HHG generation experimentation, the setup necessitated careful alignment of a Mach–Zehnder interferometer [40]. This was to produce noncollinear driving lasers that could converge at the generation medium's crossing plane. A system diagram of our novel passive approach is shown in Fig. 5. When a linearly polarized beam encounters the transmissive grating beam splitter (Boulder Nonlinear Systems), it is angularly separated into beams of opposite circular polarization with high efficiency. After reflecting from a flat mirror, the two beams pass through the splitter again, resulting in two parallel beams that are focused with a lens. The overlap of the beams at the focus results naturally from the passive architecture of the optics.

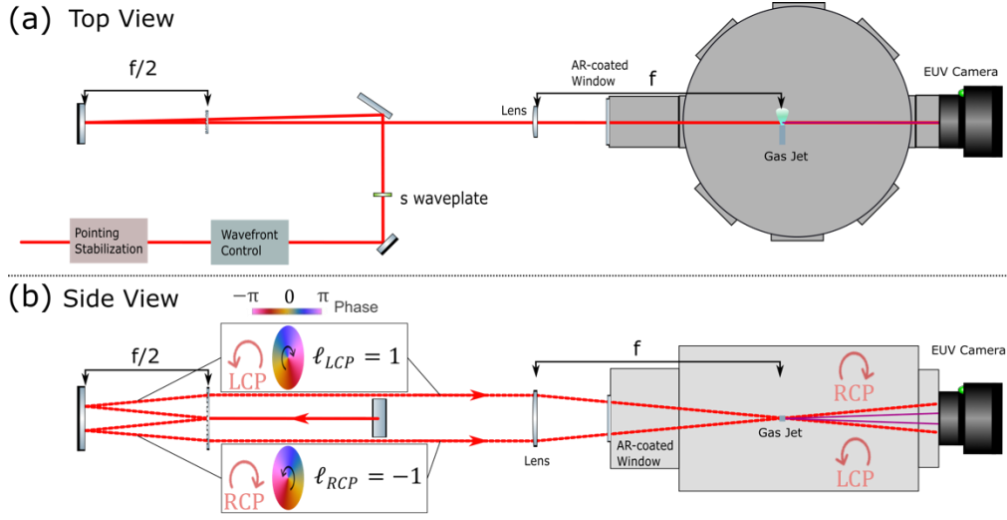

Fig. 5. Noncollinear HHG experimental setup to produce a polarization tilt-angle fork grating that drives the generation of high-order harmonics with the same topological charge. The laser beam first passes through pointing stabilization and wavefront control. The tailored beam (linearly polarized) is then sent into a grating that angularly separates the two beams of opposite circular polarization. Those beams are directed back to the grating to have parallel beams that can be focused into the vacuum chamber HHG. Panel (a) shows the top-down view of the setup, and (b) the side view where the splitting of the beams occurs.

When crossing two ultrafast pulses that are produced in a Mach-Zehnder interferometer with a lens, the pulse fronts also cross at an angle, which can result in a temporal walk-off of the two pulses that depends on the transverse position in the crossing plane [76]. The new setup presented here allows for the matching of pulse front tilts of the two beams. This is achieved by setting the effective grating-to-grating distance equal to the lens focal length. Although the crossing angle in this experiment is small, and the pulse durations are sufficiently long to make the temporal walk-off effect important, this feature of our optical arrangement will become important as this approach is extended to driving harmonics with shorter pulses.

In the present experiment, the experimental setup is further enhanced by passing the beam through an s-waveplate, fabricated by the group of Prof. P. Kazansky, to produce a vector beam upstream of the beam-splitter system. Vector beams can be expressed as a superposition of vortex beams with opposite circular polarization and OAM helicity. Consequently, when the vector beam passes through the beam splitter, the beam decomposes into circularly polarized vortex beams with opposing OAM. The use of a single s-waveplate ensures that the vector beam singularity is positioned identically on the two OAM beams, and the passive architecture of the beam-splitter system ensures the precise spatial alignment of the focal spot that will generate high-quality vortex beams. We note that in an interferometric system, separate spiral phase plates would need to be aligned along each arm which introduces extra degrees of possible error.

After generation, the resulting harmonics are either collected directly onto an EUV Camera (PIXIS-XO-1024B) or spectrally dispersed from a reflection grating. Two aluminum filters of 200 nm thickness are used before the camera sensor to block the fundamental beam from detection in either scenario. The result of the direct measurement is shown in Fig. 2d. The vacuum system also has a curved grating and Au mirror pair that can be inserted into the optical path to provide a spectrally sensitive measurement. This allows for a spectral separation on the harmonics while preserving some spatial qualities in the dispersed direction, as seen in Fig. 3d.

### *6.2 Numerical simulations*

Our theoretical simulations of HHG rely on a discrete dipole approximation model [77] that considers the macroscopic argon gas target as multiple point-like dipole sources. This is justified because the inhomogeneity of the driving beam occurs at much larger scales than the size of the atoms. The high-harmonic field from every emitter is calculated within a quantum model in the strong-field approximation [78]. Afterwards, we apply the Maxwell propagator of a point-like source to obtain the far-field beam at the detectors [77]. The total field is, thus, the coherent sum of all the contributions from the target's atoms. For the spatial distribution of the driving beams, we consider the analytical expression of Laguerre-Gaussian modes [79]. Note that this theoretical method has been successfully employed in diverse works yielding an excellent agreement against experiments [66–70,80,81].

In the results presented in this work, the gas target is modeled as a thin layer, an approximation that has been proven valid in HHG schemes where the dominant macroscopic effect is transverse phase matching [82]. In particular, for a noncollinear geometry, we have observed that the longitudinal dimension can be neglected for interaction lengths much smaller than the Rayleigh distance, $z_R = \pi w_0/\lambda$. In our calculations, we set a peak intensity of $1.72 \times 10^{14}$ W/cm$^2$, a central wavelength of 800 nm, beam waists of $w_0 = 50$ $\mu m$, and a half-crossing angle of $\theta_c = 0.85°$ to mimic the experimental conditions. The temporal profile of the laser pulse is modeled with a sine-squared envelope. The full-width duration at half-maximum intensity is chosen to be 7.7 fs, which ensures a multi-cycle pulse behavior. Note that this value is lower than the experimental pulse duration, but it has been chosen to speed up the computations. This does not modify the essential physical picture, apart from the number of pulses obtained in the attosecond vortex pulse trains (and the spectral width of each harmonic order).

**Funding.** European Research Council (851201); Ministerio de Ciencia e Innovación (PID2019-106910GB-I00, PID2022-142340NB-I00); Air Force Office of Scientific Research (FA9550-22-1-0495).

**Acknowledgments.** We would like to thank the group of Prof. P. Kazansky for the fabrication of the s-waveplate used in the experiment. The project leading to this publication has received funding from the European Research Council (ERC) under the European Union's Horizon 2020 research and innovation programme (grant agreement No 851201), and from the Air Force Office of Scientific Research (FA9550-22-1-0495). We acknowledge the computer resources at MareNostrum and the technical support provided by Barcelona Supercomputing Center (FI-2022-3-0041, FI-2023-3-0045).

**Data availability.** Data underlying the results presented in this paper are available from the authors upon reasonable request.

**Disclosures.** The authors declare no conflicts of interest.

# ATTOSECOND VORTEX PULSE TRAINS: SUPPLEMENTAL DOCUMENT

ALBA DE LAS HERAS,[1,2,4] DAVID SCHMIDT,[3,5] JULIO SAN ROMÁN,[1,2] JAVIER SERRANO,[1,2] DANIEL ADAMS,[3] LUIS PLAJA,[1,2] CHARLES G. DURFEE,[3] AND CARLOS HERNÁNDEZ-GARCÍA[1,2]

[1]*Grupo de Investigación en Aplicaciones del Láser y Fotónica, Departamento de Física Aplicada, Universidad de Salamanca, E-37008 Salamanca, Spain*
[2]*Unidad de Excelencia en Luz y Materia Estructuradas (LUMES), Universidad de Salamanca, Salamanca, Spain*
[3]*Department of Physics, Colorado School of Mines, Golden, Colorado 80401, USA*
[4]*albadelasheras@usal.es*
[5]*daschmid@mines.edu*

In this Supporting Information, we present additional figures with the numerical results of attosecond vortex pulse trains carrying higher topological charges. This shows the versatility of our technique to tailor attosecond vortex pulse trains with any topological charge. Fig. S1 displays the spatiotemporal isosurfaces of the electric field and intensity of the attosecond pulse trains arising from high-harmonic generation driven by $\ell_{RCP} = -2$ and $\ell_{LCP} = 2$. As expected, the electric field in Fig. S1(a) describes two intertwined helicoids with opposite handedness, whereas again a toroidal distribution is associated with the intensity in Fig. S1(b).

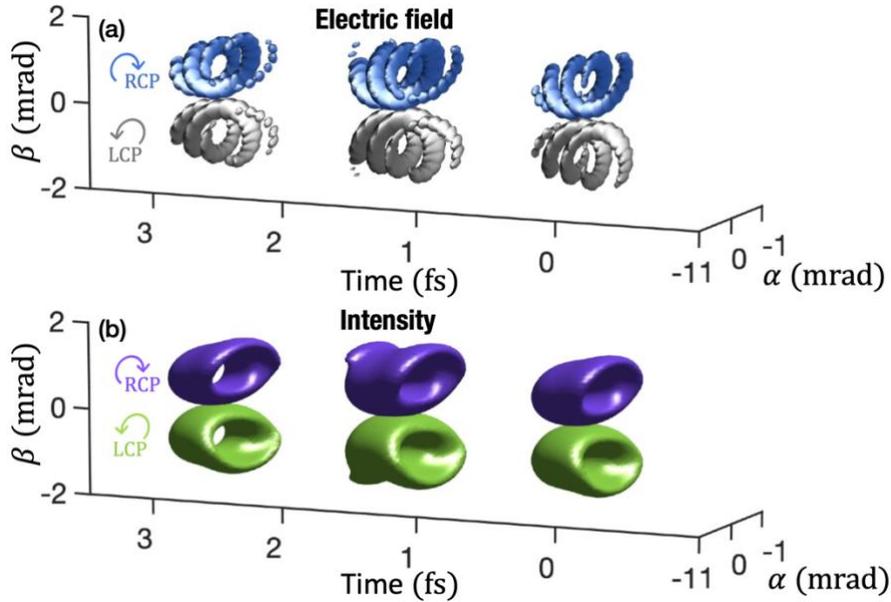

Fig. S1. (a) Electric field and (b) intensity spatiotemporal structure of the attosecond vortex pulses with topological charges $\ell_{RCP} = -2$ and $\ell_{LCP} = 2$.

In Fig. S2, we depict the same results but for $\ell_{RCP} = -3$ and $\ell_{LCP} = 3$. In this case, we distinguish three intertwined helices in the electric field structure. We note a larger divergence of the beams as the topological charge is increased. This is especially evident in the toroids of the intensity profile. We note that the two high-harmonic beams present a certain overlap. Nevertheless, the beam's separation at the far-field can be tuned by adjusting the noncollinear crossing angle and the beam's waists.

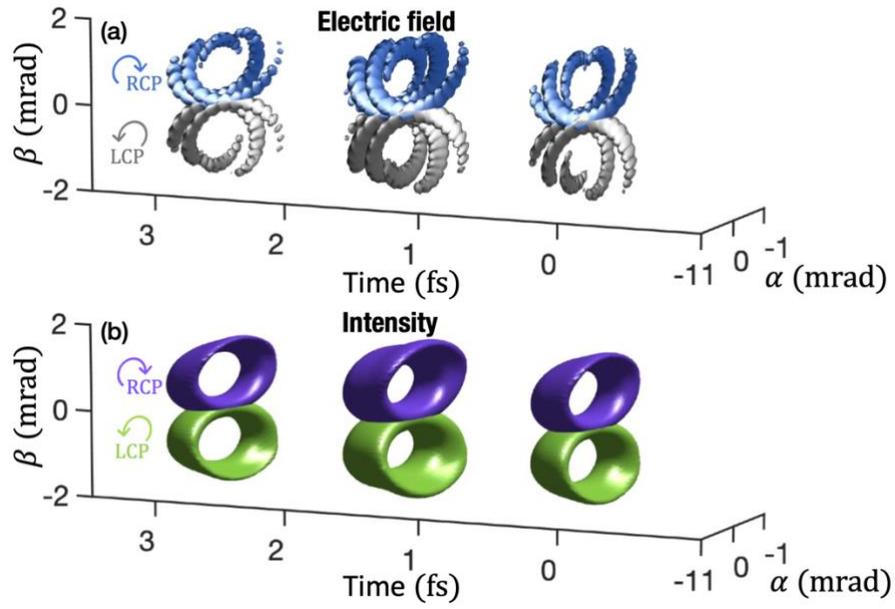

Fig. S2. (a) Electric field and (b) intensity spatiotemporal structure of the attosecond vortex pulses with topological charges $\ell_{RCP} = -3$ and $\ell_{LCP} = 3$.